\documentclass[conference]{IEEEtran}
\IEEEoverridecommandlockouts

\usepackage{cite}
\usepackage{amsmath,amssymb,amsfonts}
\usepackage{algorithmic}
\usepackage{graphicx}
\usepackage{textcomp}
\usepackage{xcolor}
\usepackage{algorithm}
\usepackage{algorithmic}
\usepackage[a4paper, total={184mm,239mm}]{geometry}
\def\BibTeX{{\rm B\kern-.05em{\sc i\kern-.025em b}\kern-.08em
		T\kern-.1667em\lower.7ex\hbox{E}\kern-.125emX}}
\begin{document}
	
	\title{Generative AI-Based Effective Malware Detection for Embedded Computing Systems 
	}

	\author{\normalsize
		\begin{tabular}[t]{ccc}
			\large Sreenitha Kasarapu & \large Sanket Shukla & \large Rakibul Hassan \\
			George Mason University & George Mason University & George Mason University\\
			

			Virginia, USA & Virginia, USA & Virginia, USA\\
			\vspace{2em}
			
			e-mail: skasarap@gmu.edu & e-mail: sshukla4@gmu.edu & e-mail: rhassa4@gmu.edu\\

			\large Avesta Sasan & \large Houman Homayoun & \large Sai Manoj Pudukotai Dinakarrao \\
			University of California Davis & University of California Davis & George Mason University\\
			California, USA & California, USA & Virginia, USA\\
			e-mail: asasan@ucdavis.edu & e-mail: hhomayoun@ucdavis.edu & e-mail: spudukot@gmu.edu\\
			\vspace{-10pt}
		\end{tabular} \vspace{-1em}}

	\maketitle
	
	\begin{abstract}
		One of the pivotal security threats for the embedded computing systems is malicious software \textit{a.k.a} malware. With efficiency and efficacy, Machine Learning (ML) has been widely adopted for malware detection in recent times. Despite being efficient, the existing techniques require a tremendous number of benign and malware samples for training and modeling an efficient malware detector. Furthermore, such constraints limit the detection of emerging malware samples due to the lack of sufficient malware samples required for efficient training. To address such concerns, we introduce a code-aware data generative AI technique that generates multiple mutated samples of the limitedly seen malware by the devices.
		Loss minimization ensures that the generated samples closely mimic the limitedly seen malware and mitigate the impractical samples. Such developed malware is further incorporated into the training set to formulate the model that can efficiently detect the emerging malware despite having limited exposure. The experimental results demonstrates that the proposed technique achieves an accuracy of 90\%, which is approximately 9\% higher while training classifiers with only limitedly available training data.

	\end{abstract}
	

\section{Introduction}

With the technical developments in hardware architecture and embedded systems, IoT applications have procured enormous interest in the past few decades \cite{iot-1}. Currently, manufacturers are experiencing an immense desire to automate user applications and produce interactive software systems such as smart homes, digital monitoring, and smart grid. These systems handle vast amounts of user data daily and are vulnerable to security threats \cite{Abbas2016BigDI} due to malicious software or malware. There have been more than 5 billion malware attacks worldwide, only in the year 2020 \cite{stat_1}.

Despite the advanced anti-malware software, malware attacks increase because of the newer emerging-malware signatures each year. Adversaries generate millions of new signatures of malware each year \cite{stat_2} to steal valuable information for financial benefit and stay undetectable. The tremendous rise in the volume of malware attacks poses a severe threat to hardware security \cite{cyber-risk-2}. Thus it is vital to detect the malware, as it can exploit confidential user information, leading to a substandard user experience. Realizing the threat caused by malware in terms of stolen information, access to sensitive information like passwords, and billions of revenue loss, severe measures are being taken to abate malware escalation. Static and dynamic analysis \cite{sta_dy},\cite{prove_analysis} are employed to detect malware, but these techniques are time-consuming and not efficient to identify obscure malware families. Use of Machine Learning (ML) for malware detection is seen as an efficient technique \cite{img_process,img_vis}.

Works such as \cite{img_vis, img_process} obtains the binaries of benign and malware files, transform them into 
images, fed to the ML classifiers for malware detection. Use of such technique is efficient compared 
to other techniques due to their prime ability to learn image features. 
However, one of the main challenges with adopting such a technique is the requirement of massive amounts of training samples. 
With the exponential increase in the generation of newer malware families each year, it is complex to obtain a sufficient number of malware samples for each new malware class. 

Furthermore, malware developers implement code obfuscation, metamorphism, and polymorphism \cite{obfuscation,morphism,sanket_abhijitt_date2021,sanket_cases_2019,sanket_dac_2021,sanket_date_2023,sanket_fl,sanket_glsvlsi_2022,sanket_iccd_2022,sanket_icmla_2019,sanket_isqed_2024,sanket_rram_glsvlsi_21,sanket_sreenitha_metrocad,sanket_sreenitha_tcad} to mutate malware binary executables. The mutated malware resembles the functionality of a standard application, thereby deceiving the malware detection mechanism. Code obfuscation is a technique where specific parts of the code in malware binary files are encapsulated, masking the malware's behavioral patterns without affecting its functionality. Adversaries use this to evade detection and prolong their presence in the embedded systems to exploit their security.
A new strategy in masking malware's identity is stealthy malware \cite{stealthy}, where malware is incorporated into benign applications using random obfuscation techniques. Malware attackers also sneak malware into benign application files to hide its vulnerability. This stealthy malware does not exhibit many malware characteristics. Some stealthy malware can take a long time before revealing any malicious activity and adversely impact system security.

In this work, we address all the issues mentioned above. We propose a code-aware data generation technique that can generate mutated training samples and capture the features of actual samples. The generated images mimic the limitedly seen malware and resolve the demand for comprehensive data acquisition.  
Our proposed paradigm detects malware that impersonates or abuse benign applications. We chose code obfuscated and stealthy malware that is most complex to detect using traditional static and dynamic malware detection techniques.


The novel contributions of this work can be outlined in a three-fold manner: 
\vspace{0.0003em}
\begin{itemize}
	
	\item Introduces a code-aware generative-AI architecture for increasing the training dataset.
	\item Loss minimization is employed to ensure, that the generated data, captures the code patterns of limitedly seen complex malware and its functionality is preserved.
	\item Few-shot learning is used to classify complex stealthy malware and code obfuscated malware efficiently.

\end{itemize}


The proposed approach can efficiently classify complex malware by using only a limited number of samples. The experimental results demonstrate that the proposed technique can achieve up to $89.52\%$ accuracy, which is 7\% higher compared to models trained only on limited samples. 

The rest of the paper is organized as follows: Section \ref{related_work} describes the related work and their shortcomings and comparison with the proposed model. Section \ref{prob_form} describes the problem formulation. Section \ref{prop_tech} describes the proposed architecture, which assists with data-aware sample generation for efficient malware detection, even with limited exposure of training samples. The experimental evaluation of the proposed model and comparison with various ML architectures is illustrated in Section \ref{exp_results}, and then we conclude in Section \ref{conclusion}.

\section{Related Work}

\label{related_work}

Traditionally malware detection is carried out using static and dynamic analysis. Static analysis \cite{sta_dy} is performed in a non-runtime environment by examining the internal structure of malware binaries and not by actually executing the binary executable files. Static analysis is not an efficient approach for malware detection but serves as a quick testing tool \cite{static_limits}. In dynamic analysis, the binary applications are inspected as malware, or benign by executing in a harmless, isolated environment \cite{sta_dy,dynamic}. Unlike static analysis, dynamic analysis is a functionality test. Although dynamic analysis can handle complex malware, it can only expose malware that exhibits malware properties \cite{dynamic}. Dynamic analysis is not efficient in detecting hidden malware code blocks which restrain from getting executed, and it is a time-consuming process. 

Later \cite{nataraj} introduced a technique for malware detection using image processing where binary applications are converted into grayscale images. The generated images have identical patterns because of the executable file structural distributions. The paper used the K-Nearest Neighbour ML algorithm for the classification of malware images. Other approaches \cite{img_vis, img_process} include image visualization and classification using machine learning algorithms such as SVM. However, these approaches don't address the problem of classifying newer complex malware that is code obfuscated, polymorphed, etc. Neural networks such as ANNs are used extensively to solve the problem \cite{img_process}, as neurons can capture the features of the images accurately than other machine learning algorithms. But, the fully connected layers of artificial neural networks tend to exhaust computational resources. 
In \cite{cnn_detect, sreenitha_covid, sreenitha_mdpi, sreenitha_sanket_aspdac, sreenitha_sanket_glsvlsi, sreenitha_sathwika_vlsid, sanket_sreenitha_tcad, sanket_sreenitha_metrocad, raghul_iscas, raghul_SaravananICICNIS'21, raghul_RS2020, raghul_SaravananICDSMLA'19, raghul_trng2020, Raghul2019} authors used Convolutional neural networks for malware confinement, as they are popular for their ability to efficiently handle image data through feature extraction by Convolutional 2D layers and using Maxpooling 2D layers to downsample the input parameters, thus, reducing the computational resources. The drawback here is, they need to be trained with large amounts of training data to perform classification efficiently.

We, therefore, need an efficient model which can address all the concerns mentioned above. Also, the model must efficiently classify code obfuscated and stealthy malware without the need for a vast training dataset. In this paper, we address this by developing a code-aware generator that can generate realistic images. These generated images can replicate features of various malware families, solving the problem of training the CNN for efficient malware detection with limited samples.

\section{Problem Formulation}
\label{prob_form}
The motivation for the proposed malware detection is the limited available malware data and the need for vast amounts of malware samples to train a classification model for malware detection. In the real world, with the constantly evolving malware families and the newer complex malware introduced each day, it is impossible to use a single classification model to detect various malware signatures. And, it is also impractical to update the training model each time a new malware version is released and available. Moreover, it is not easy to collect enough data samples of each type of malware and perform efficient classification. Especially with the malware disguising themselves as benign with code obfuscation and stealthy malware methods, the stakes are high to build a classification model with limited training data. They can classify malware that disguises its malware functionalities. So in this work, we try to address the problem of constructing a model that can efficiently detect malware based on limited malware data. The problem is constructed  as follows:

\vspace{0.001 em}
\begin{equation}
	\centering
	D = \{B+M+O_m+S_m\}
	\vspace{0.0001em}
	\label{eq1}
\end{equation}

Given a dataset $D$ containing $n$ samples $(D_1, D_2, .., D_n)$ comprising of four classes benign $B$, traditional malware $M$, complex random obfuscated malware $O_m$ and stealthy malware $S_m$ as shown in equation \ref{eq1}
Dataset $D$ has the required number of samples $n$ to train a Machine Learning model for efficient malware detection. Also, the complex data avaliable is represented by dataset $D_l^x$, and has much less samples, represented by $x$ and its value is less than or equal to $\nabla$\% of samples $n$, from dataset $D$ as shown in equation \ref{eq2}.$\nabla$ is a numerical value and $\nabla$\% $n$ are a limited of samples available for training. 

\vspace{0.0001 em}
\begin{equation}
	\centering
	D_l^x \subset D^n; \hskip 0.5em  \forall \hskip 0.5em \{x <= \nabla\% n\}
	\vspace{0.1em}
	\label{eq2}
\end{equation}


Despite the limitedly available complex data for training an ML model, one should devise a technique that can efficiently classify between these limitedly seen malware data and benign samples.

\vspace{0.0001 em}
\begin{equation}
	\centering
	\textbf{C:}(D_l) \Rightarrow (B, M, O_m, S_m)
	\vspace{0.1em}
	\label{eq3}
\end{equation}

As shown in \ref{eq3}, given a dataset $D_l$ containing limited number of complex malware and benign samples, a classifier $\textbf{C:}$ must be built which can efficiently classify between limitedly seen classes, such as, benign $B$, traditional malware $M$, complex random obfuscated malware $O_m$ and stealthy malware $S_m$.







\section{Proposed Technique}
\label{prop_tech}
\subsection{Overview of the Proposed Technique}\label{AA}
\begin{figure*}[htbp]
	\centering
	\includegraphics[width=7in]{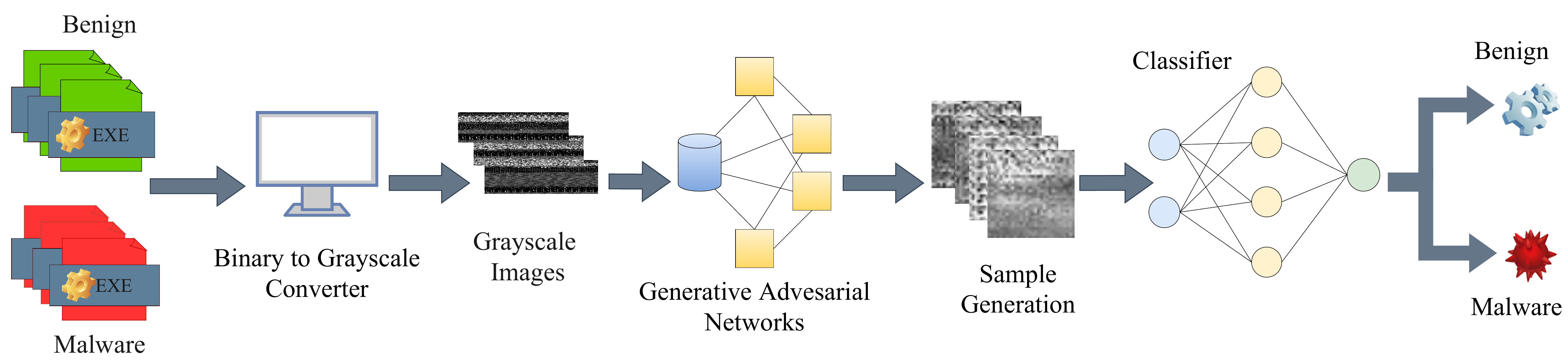}
	\caption{Overview of Proposed Code-Aware Sample Generator}
	\label{fig:generator}
	
\end{figure*}

The overview of the proposed technique is shown in Figure \ref{fig:generator}. The solution initiates by collecting benign and malware application binary files as input. Among these, malware binaries are a severe threat to the system. The malware files collected belong to trojan, backdoor, worm, virus, rootkit families, and code obfuscated malware versions. Stealthy malware is generated using code obfuscated techniques from the available benign and malware samples, while obfuscated malware is generated using the random obfuscation technique. These files are processed as an input to the proposed technique.

\begin{algorithm}
	\caption{Pseudo-Code for proposed technique}
	\begin{algorithmic} 
		\label{algo1}
		\STATE  \hskip -1em {\textbf{Require}}: $(B_{exe})$ (Benign application files), $(M_{exe})$ (Malware application files), $B$ (Benign grayscale images), $M$ (Malware grayscale images), $O_m$ (Random obfuscated malware), $S_m$ (Stealthy malware),$D^n_w$ (Whole Dataset with n samples), $D_l^x$ (Dataset with limited data version containing x samples)
		
		\STATE  \hskip -1em {\textbf{Input}}: $ B_{f} = \{(B
		_{exe})^a, (M_{exe})^b\}; $
		
		\STATE $\forall \hskip 0.5em m \leftarrow (1, 2,3,...a),$ $n \leftarrow (1,2,3,...b) $
		
		\ENSURE $ B_{f} \Rightarrow Grayscale\_Images $
		
		\STATE $ \{B =(B_1, B_2, B_3,....B_a) ;  M = (M_1, M_2,M_3,....M_b) \} \Rightarrow Input\_images$
		
		
		\STATE $\textbf{stealthy\_generator}(B_{exe}, M_{exe}) \Rightarrow S_m$
		\STATE $\textbf{random\_obfuscator}(M) \Rightarrow O_m$
		
		\STATE $D^n_w = \{B + M + O_m + S_m\} \Rightarrow D_{l}^x $
		
		\STATE  $ \textbf{CAD\_generator}(D_l^x) \Rightarrow D_m^y  $
		
		\STATE $ Proposed =  \{D_l^x \cup D_m^y\}$
		
		\STATE $ \textbf{Classifier}(D), \textbf{Classifier}(Proposed) \Rightarrow (B, M, O_m, S_m)$
	\end{algorithmic}
\end{algorithm}

Algorithm \ref{algo1} overviews the pseudo-code for the proposed technique. Even though Convolutional Neural networks are one of the top-performing classification models, data must be an image to employ them for the classification problem. So, the application files \verb|.exe| are converted into grayscale images for training the CNN model. The $B_f$ folder contains benign, and malware application files represented as $B_{exe}$ and $M_{exe}$, containing $a$ and $b$ number of samples respectively. Random Obfuscated malware $O_m$ is generated using malware ($M$) image samples, which are passed through the Random obfuscation generation function represented as $ {random\_obfuscator}(M) $. Stealthy malware $S_m$ samples are generated using the combination of benign and malware binary samples, which are passed through the stealthy generation function represented as $ {stealthy\_generator}(B_{exe}, M_{exe}) $. The benign and malware samples are combined iteratively to generate $S_m$ samples.

The dataset $D_w$ is the entire pool containing $n$ number of samples comprising benign $B$, malware$M$, obfuscated malware $O_m$, and stealthy malware $S_m$ classes. We create limited data samples $D_l^x$  from the dataset $D_w$ containing only $\nabla$\% of the original samples $n$, randomly picked from each class of the dataset $D$. The number of samples in $D_l$ is represented by $x$. By utilizing dataset $D_l$, generative adversarial networks are trained to generate code-aware mutated fake images, represented with the function $ CAD\_generator(D_l^x$). The mutated images are stored in dataset $D_m$ and limited data version dataset $D_l$ are combined to create proposed dataset $Proposed$ with $k$ samples. Multi-class classification is performed on both datasets $D_w$, $Proposed$ using the same CNN architecture.

\subsection{Data Generation}\label{AA}

The steps shown in the figure \ref{fig:generator} are followed to generate input data for classification. The generator takes the binary applications file as input and converts them into a raw binary bitstream. This binary bitstream is then converted to an 8-bit vector. Each 8-bit vector containing the binary values is taken as a byte, representing different image pixels. As the output image is grayscale, 0 represents white, and 255 represents black color in the grayscale image, the rest of the pixels are intermediate shades of gray. The output size of the image is based on the application file given as input to the generator model. The image's width is fixed to 256, and the height is variable based on the size of the application file arranged as an image.

\subsubsection{Stealthy Malware Generation}

Stealthy malware is created by obfuscating malware into benign. It hides the malware functionality in the benign processes, as it has both malware and benign elements. A code obfuscator is built to generate stealthy malware using input $M_{exe} $ and $B_{exe}$ files.

\begin{algorithm}
	\caption{Stealthy Malware Generation Algorithm}
	\label{algo2}
	\begin{algorithmic} 
		
		\STATE  \hskip -1em {\textbf{Require}}: $(B_{exe})$ (Benign application files), $(M_{exe})$ (Malware application files), $\xi$ (Code Obfuscation function), $O_m$ (Random obfuscated malware)
		\STATE  \hskip -1em {\textbf{Input}}: $B_{exe}, M_{exe}$
		
		
		\STATE \textbf{define} $stealthy\_generator(B_{exe}, M_{exe})$:
		
		\STATE \hskip 1em \textbf{for} {$ i \leftarrow range(len(a))$}: \textbf{do} 
		
		\STATE \hskip 2em \textbf{for} {$j \leftarrow range(len(b))$}: \textbf{do}

		\STATE \hskip 3em $\xi((B_{exe})^a\oplus (M_{exe})^b) \Rightarrow S_m$

		\STATE \hskip 2em \textbf{end for}
		
		\STATE \hskip 1em \textbf{end for}$ $
		\STATE \hskip 1em \textbf{return} $S_m$
		
	\end{algorithmic}
\end{algorithm}

Algorithm \ref{algo2} represents the steps to be followed to generate stealthy malware. The random obfuscation function represented as $\xi((B_{exe})^a\oplus (M_{exe})^b)$ takes in the benign and malware binary files as input, breaks the code blocks in both files, converts certain parts of the code unreadable and shuffles the code structures from both files to create a single binary file. Code obfuscation changes the code structure of binary files, but it still retains the malware functionality. Identifying code obfuscated malware is complex as it contains the majority of traces of benign features. The generation is iterated to cover all the possible combinations between $a$ number of benign and $b$ number malware files.

\subsubsection{Random Obfuscated Malware Generation}
Random obfuscation is a technique that can mask the malware functionality in a sample. We generate random obfuscated malware samples as a part of the dataset. The process to randomly obfuscate a malware image is iterative, wherein each iteration, a random obfuscator considers a different input image, divides it into segments, and randomly re-arranges it. 
The shape of each grayscale image is dependent on the size of the input binary file used to create it. Each input image is fixed to a width of 256, and the height is variable. Further, the grayscale image is divided into segments and arranged randomly to produce an obfuscated malware image. However, it is essential to note that the random obfuscation retains the malicious functionality of the obfuscated malware image.

\begin{algorithm}
	\caption{Random Obfuscated Malware Generation}
	\label{algo3}
	\begin{algorithmic} 
		\STATE  \hskip -1em {\textbf{Require}}: $M$ (Malware grayscale images), $S_m$ (Generated Stealthy malware)
		\STATE  \hskip -1em {\textbf{Input}}: $M = (M_1, M_2, M_3,....M_n)$, $\forall \hskip 0.5em n \leftarrow (1, 2,3,...n)$
		\ENSURE $ w, h \leftarrow M_{img}.size $

		
		\STATE \textbf{define} $random\_obfuscator(M)$:
		\STATE \hskip 1em \textbf{for} {$ M \leftarrow range(0,n)$}: \textbf{do}

		\STATE \hskip 2em \textbf{for} {$j \leftarrow range(0,w,width)$}: \textbf{do}
		
		\STATE \hskip 3em \textbf{for} {$j \leftarrow range(0,w,width)$}: \textbf{do}

		\STATE \hskip 4em $box \leftarrow(j, i, j+width, i+height)$
		\STATE \hskip 4em $ list \leftarrow M_{img}.crop(box)$ ; $ rand(list) $
		\STATE \hskip 4em $ O_m \leftarrow vstack(list) $
		
		\STATE \hskip 3em \textbf{end for} 
		
		\STATE \hskip 2em\textbf{end for}
		
		\STATE \hskip 1em\textbf{end for}
		
		\STATE \textbf{return} $O_m$
	\end{algorithmic}
\end{algorithm}

The algorithm \ref{algo3} describes the steps involved in randomly obfuscating malware samples. 
Malware images are taken as input represented as $M$. As the width of these images is fixed to 256 and height is variable to find the height of the input images $M_{img}.size$ function is used. The images are segmented vertically, with the height of the segments fixed to 10, and the width is constant to 256. The malware $M$ images are cropped using the box function coordinates, creating segments for each size $(256,10)$. Using these coordinates, a box function is constructed and iterated to cover the whole image. The segments from $M$ images are then added to a $list$ as represented in the algorithm. The segments are shuffled and randomly reconstructed by vertically stacking the segments $vstack(list)$, to create a randomly obfuscated malware image $O_m$. Nevertheless, it still has all of its components and unchanged functionality with the added benefit of its ability to mask malware components in itself and making it hard to detect.

\subsubsection{Code-Aware Data Generation}
Code-aware data generation is a novel approach to generate mutated data for malware detection. Code-aware data generation is a techniques where pseudo data is generated using malware binary code. The malware binaries are converted into images, these images capture the code patterns, representing the malware functionalities and are then given as input to generate more samples. The generated data interpert the malware behaviour, making them malware code-aware samples. The generated images can increase the efficiency of the classification model by participating in the training of limitedly available complex malware data. The generated images are loss-controlled, so they are good at capturing limited malware data features. We use generative adversarial networks(GANs) to perform this task. GANs consist of two neural networks known as generator and discriminator. Generators consider a random uniform distribution as a reference to generate new data points. The generator works to generate fake images which are similar to the input data. The discriminator's work is to classify the fake images produced from the generator to the real images. The generator and discriminator have the opposite functionality. The generator tries to generate fake images, which the discriminator can't classify as fake. The discriminator tries to learn and make itself able to discriminate more and more fake images. They compete with each other to invalidate the other's function and improve themselves in the process of learning. The generator and discriminator try to control their loss function. The generator learns to generate images much similar to the real images the discriminator learns to classify them better. With enough training steps, the generator generates realistic images and can be used as a data generator.

\begin{algorithm}
	\caption{Code-Aware Data Generation Algorithm}
	\label{algo4}
	\begin{algorithmic}

		\STATE  \hskip -1em {\textbf{Require}}: $D_{l}^x$ (Dataset with limited data version containing x samples), $O_m$ (Random obfuscated malware), $B$ (Benign grayscale images), $S_m$ (Generated Stealthy malware)
		\STATE  \hskip -1em {\textbf{Input}}: $ D_{l}^x = \{B+ M +O_m+ S_m\} $

		\STATE \textbf{for} {$ X \leftarrow D_{l}$}: \textbf{do} 
		
		\STATE \hskip 1em \textbf{define} CAD\_generator(X):
		
		\STATE \hskip 2em \textbf{for} {$epoch \leftarrow range(1000)$}: \textbf{do}

		\STATE \hskip 3em $G\_model = define\_generator()$
		\STATE \hskip 3em $D\_model = define\_discriminator()$
		\STATE \hskip 3em $n \leftarrow noise\_vector(256, None)$ 
		\STATE \hskip 3em $X\_{fake} \leftarrow G\_model(n) $
		
		
		
		\STATE \hskip 2em\textbf{end for}
		\STATE \hskip 2em $D_m \leftarrow G\_model.predict(vector)$
		\STATE \hskip 2em \textbf{return} $D_m$

		\STATE \textbf{end for}
	\end{algorithmic}
\end{algorithm}

The algorithm \ref{algo4} takes in the limited version dataset $D_l$ as input. For each class in the dataset, the CAD\_generator trains a generator and a discriminator. We train our GAN for 1000 epochs, enough times to minimize the loss and generate images similar to training data. As represented in the algorithm, the generator model is described as $G\_model$, and the discriminator model is described as $D\_model$. $G\_model$ and $D\_model$ are convolutional neural networks where $G\_model$ is trained to generate an image when a latent space is given as input. As represented in the algorithm, when a latent noise of size $(256, None)$ is given as input, it generates an image of size $(32,32)$. The $D\_model$ tries to classify the generated fake image X\_{fake}. A loss function is generated for $D\_model$ and $G\_model$. To decrease the gradient loss, the generator learn to generate better fake images $X\_{fake}$, and the discriminator keeps on learning to classify them. After 1000 epochs, the generator model learns enough to be able to generate realistic fake images. So vectors of latent spaces are created to generate mutated data by using the $model.predict()$ function, they are represented using dataset $D_m$.


\subsection{Classification}
Convolutional Neural networks are used for the classification task. CNN's are a type of neural network build to mimic the human visual cortex. The receptive field of a CNN comprises sub-regions layered over each other to cover the entire visible area. The ability of CNN to extracts features from an image has made it prominent to be used in visual recognition tasks such as image classification or object detection. Convolutional Neural Networks have proven to perform exceptionally well with understanding the image features. Even though the input data have similarities, CNNs tend to classify them efficiently.

\begin{figure}[htbp]\vspace{-0.75em}
	{\includegraphics[width=1\columnwidth]{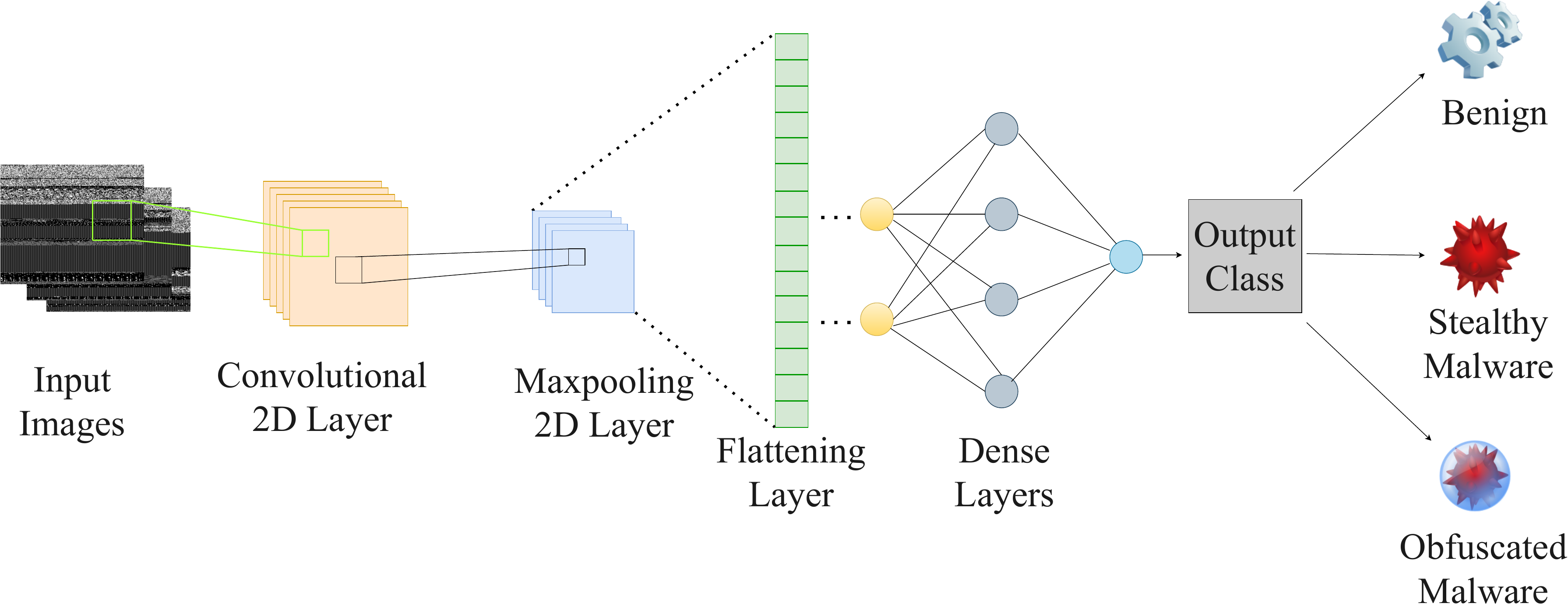}}
	\caption{ Classifier trained for Malware Detection
	}
	\label{fig:classifier}\vspace{-0.75em}
\end{figure}

The dataset has three classes benign, stealthy malware, and code obfuscated malware. A CNN model is built to classify the data. The flow of the classification task is as represented in the figure \ref{fig:classifier}. Image data is not labeled instead read using the flow from directory Keras function. The data is divided into 75\% training set, 15\% test set, and 10\% validation set. Feature extraction is done on the images using the convolutional 2D layers of the CNN architecture with (3 x 3) filters, followed by the max-pooling 2D layers.  The data is then flattened to pass it to the dense layers. All the layers of the CNN architecture are built using a \textit{`relu'} activation function, except the dense output layer, which is equipped with a \textit{`softmax'} activation to produce probabilities for each class. The image is classified into the class with the highest probability. We train two different CNNs with the same architecture for two datasets, the full dataset $D$ and the lite version dataset $D_l$. The performance of the CNN model on these datasets and comparisons with other ML models are analyzed in Section \ref{exp_results}.

\vspace{-0.5em}
\section{Experimental Results}

\label{exp_results}

\subsection{Experimental Setup}

The proposed methodology is implemented on an Intel core Nvidia GeForce GTX 1650 GPU with 16GB RAM. We have obtained malware applications from VirusTotal \cite{rvirus} with 12500 malware samples that encompass 5 malware classes: backdoor, rootkit, trojan, virus, and worm. We collected about 13700 benign application files, which are harmless to work with. From malware families backdoor, trojan, virus, and worm, 4800 random obfuscated malware samples are generated. About 6000 stealthy malware samples are generated.
The limited version of whole dataset $D_w$ with less than 30\% of data samples, is represented as $D_l$. 
Few-shot is the case where we train the CNN model with dataset $D_l$ and the synthetic data generated using Code-Aware Data generation technique. We test the CNN on rootkit data, to understand how the proposed techniques generalizes on new data.  



\subsection{Simulation Results}

\label{results}

\begin{figure}[htbp]
		\centerline{\includegraphics[width=0.7\columnwidth]{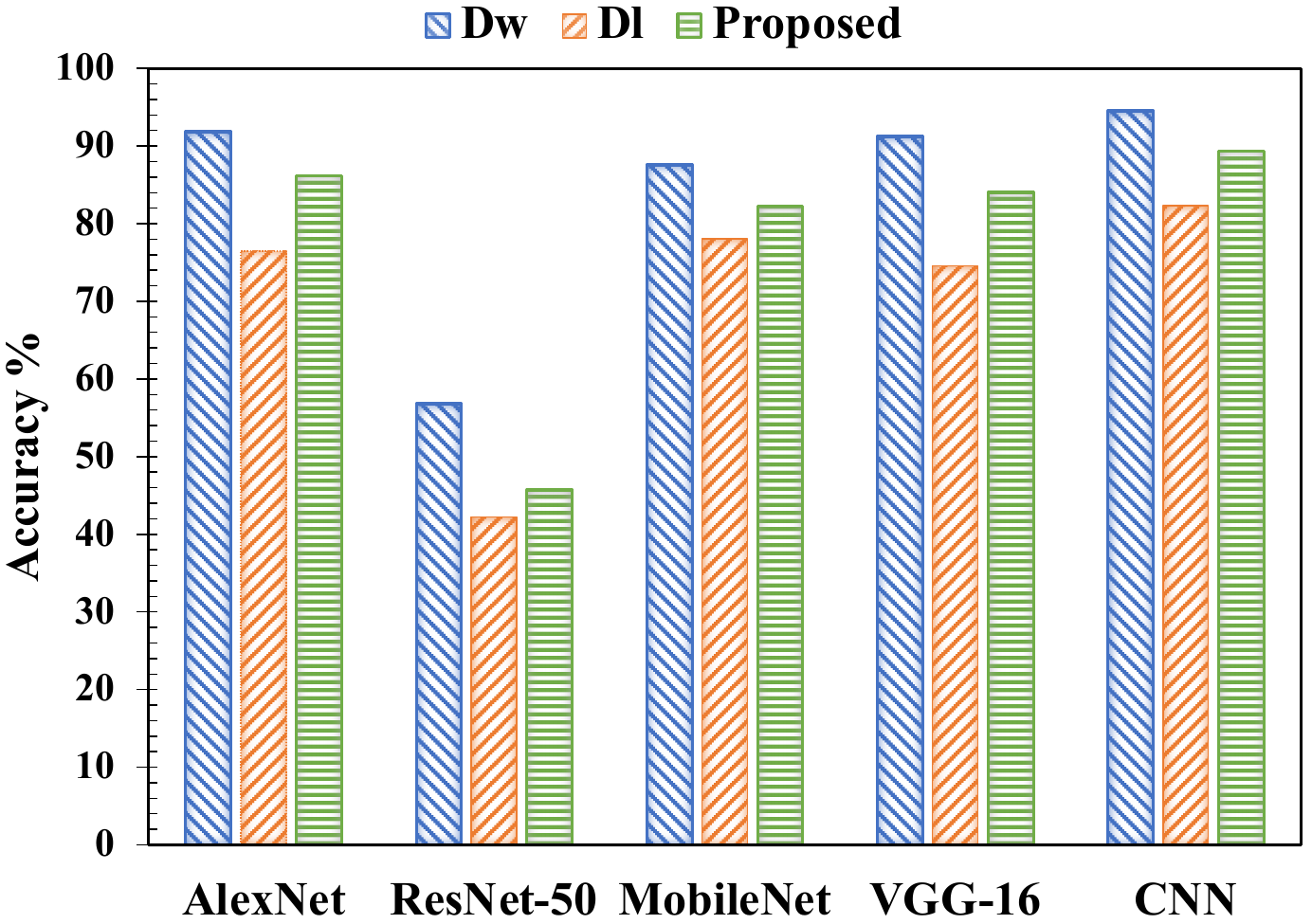}}
		\caption{ Performance analysis of various ML algorithms
		}
		\label{fig:results_graph}\vspace{-0.25em}
	\end{figure}

CNNs are trained using the three different datasets $D_w$, $D_l$, Few-shot, and the classification metrics are compared with other Machine learning models such as AlexNet, ResNet-50, MobileNet, and VGG-16. As shown in table \ref{tab2} we evaluate and compare the test accuracy, precision, recall, and F1-score of these ML models. We observed that the CNN model performs best compared to all the other classifiers in all three data setups. CNN attains the highest accuracy of $94.53\%$ when it is trained with the whole dataset $D_w$, followed by pre-trained Alexnet and VGG-16 models in the whole data setup. It is observed that the Resnet-50 model has the least performance in all the data setups. We can observe an accuracy drop of about $10\%$ in all the models when trained with only $30\%$ of the initial data, i.e., dataset limited $D_l$. The model with the highest accuracy, when trained with the $D_l$ dataset is CNN with $82.32\%$, which is low for a neural network and depicts the inefficiency in classification.
\begin{table}[htbp]
	\caption{Performance comparison of proposed model with other datasets on different ML algorithms}
	\vspace{-1em}
	\begin{center}
		\scalebox{0.8}
		{
			\begin{tabular}{|c|c|c|c|c|c|}
				\hline
				\textbf{Model} & \textbf{Dataset} & \textbf{Accuracy} & \textbf{Precision} & \textbf{Recall} & \textbf{F1-score} \\
				&  & \textbf{(\%)} & \textbf{(\%)} & \textbf{(\%)} & \textbf{(\%)} 
				\\
				\hline
				& $D_w$ & 91.84  & 91.62 & 91.58 & 91.70 \\
				\cline{2-6}
				AlexNet & $D_l$ & 76.48  & 76.45 & 76.46 & 76.45 \\ 
				\cline{2-6}
				& Few-shot & 86.15 & 86.23 & 86.28 & 86.32 \\ 
				
				\hline
				& $D_w$ & 56.79  & 56.82 & 57.00 & 56.82 \\
				\cline{2-6}
				ResNet-50 & $D_l$ & 42.25  & 42.00 & 42.25 & 42.25 \\
				\cline{2-6}
				& Few-shot & 45.73 & 45.74 & 45.78 & 44.93 \\ 
				
				\hline
				& $D_w$ & 87.58  & 87.32 & 87.45 & 87.41 \\
				\cline{2-6}
				MobileNet & $D_l$ & 78.05  & 78.03 & 78.05 & 78.03 \\
				\cline{2-6}
				& Few-shot &  82.25  & 82.05 & 82.2 & 82.15\\
				\hline
				
				& $D_w$ & 91.25  & 91.10 & 91.30 & 91.20 \\
				\cline{2-6}
				VGG-16 & $D_l$ & 74.60 & 74.30 & 74.40 & 74.40 \\
				\cline{2-6}
				& Few-shot &  84.05  & 84.00 & 83.00 & 83.00 \\
				
				\hline
				
				& $D_w$ & 94.52  & 94.18 & 94.40 & 94.15 \\
				\cline{2-6}
				CNN & $D_l$ & 82.32  & 82.03 & 82.21 & 82.18\\
				\cline{2-6}
				& Few-shot&  \textbf{89.35}  & \textbf{89.35} & \textbf{89.35} & \textbf{89.35} \\
				
				\hline
				
				DNN \cite{basnet2021ransomware} & Few-shot & - &76.6 & 74.2 & 75.3\\
				
				\hline
				RNN \cite{basnet2021ransomware} & Few-shot & - & 77.8 & 75.4 & 76.1\\
				
				\hline
				VGG-16 \cite{yue2017imbalanced} & Few-shot & - &77.1 & 74.6 & 75.2\\
				
				\hline
				Inception V3 \cite{chen2018deep} & Few-shot & - &78.6 & 74.8 & 76.3\\
				
				\hline
				Xception \cite{lo2019xception} & Few-shot & - &77.8 & 73.4 & 76.5\\
				
				\hline
				SNN \cite{SNN_ZHU2022102691} & Few-shot & - &80.3 & 82.9 & 81.8\\
				
				\hline
			\end{tabular}
		}
		\label{tab2}
	\end{center}
\end{table}
When ML models are trained with the Few-shot dataset we can see a subsequential increase in classification metrics. We can observe a performance increase in the CNN model, with its accuracy increased to $89.35\%$. More than $7\%$ of accuracy boosting can be observed compared to $D_l$, in all models except ResNet-50 with only $3\%$ accuracy boost. As shown in graph \ref{fig:results_graph}, models trained on Few-shot are performing better than models trained on $D_l$ and managed to suffer only $5\%$ of accuracy drop compared to models trained on $D_w$. This is efficient because they too have only $30\%$ of real samples as dataset $D_l$. The mutated samples in dataset Few-shot are accounting for the performance improvement compared to the dataset $D_l$. Thus, making the proposed dataset Few-shot a better option, in case of limitedly available data rather than directly using limited data $D_l$ to train the classification model. 
Compared to existing few-shot learning techniques such as \cite{basnet2021ransomware, yue2017imbalanced, chen2018deep, lo2019xception}, the F-1 score is improved by 10\% in the proposed CAD-FSL technique. Compared to the current state-of-the-art ransomware classification, based on few-shot learning (for an imbalanced dataset) \cite{SNN_ZHU2022102691}, our proposed technique has improved the F-1 score by about 8\%. 



\begin{table}[htbp]
	\caption{Quantitative measure with existing works}
	\vspace{-1em}
	\begin{center}
		\scalebox{0.75}
		{
			\begin{tabular}{|c|c|c|c|c|}
				\hline
				\textbf{Samples } &  \multicolumn{2}{c|}{\textbf{Existing works}} & \multicolumn{2}{c|}{ \textbf{Proposed}} \\
				\cline{2-5}
				
				& \textbf{Input data} & \textbf{Accuracy} & \textbf{Input data} & \textbf{Accuracy}\\
				\hline
				Malware & 14733 & 97.89\% & 700 & 92.01\%  \\
				\hline
				Ofuscated & 1172 & 96.31 \% & 340 & 89.14\% \\
				\hline
				Stealthy & 1000 & 90.80\% & 400 & 85.23\%  \\
				\hline
				
					
				\end{tabular}
			}
			\label{tab3}
		\end{center}
	\end{table}

	Table \ref{tab3}, has the quantitative analysis between existing works and proposed works. Due to the complex nature of training data, there are only limited examples of these data classifications. We considered about 30\% or fewer training samples of which existing works were used and achieved results with a slight accuracy decay. The existing work for stealthy malware \cite{stealthy_acc}, achieved 90.08\% accuracy, we can suffer only a 5\% decrease in that. The existing work for malware \cite{obfus_acc}, has an accuracy of 97.89\% and we have a 6\% decrease in that. The existing work for obfuscated malware \cite{obf_result}, has an accuracy of 96.31\% and we have a 6\% decrease in that. Considering the number of samples used for training, the proposed technique serves as the best option when only a limited amount of data is available.

	\subsection{ASIC Implementation of Proposed Technique for different Classifiers}
	
	We conducted a comprehensive hardware implementation of the classifiers embedded into the proposed technique on ASIC. All the experiments are implemented on a Broadcom BCM2711, quad-core Cortex-A72 (ARM v8) 64-bit, 28 nm SoC running at 1.5 GHz. The power, area, and energy values are reported at 100 MHz. We used Design Compiler Graphical by Synopsys to obtain the area for the models. Power consumption is obtained using Synopsys Primetime PX. The post-layout area, power, and energy are summarized in Table \ref{tab:asic_res}. Among all the classifiers, AlexNet consumes the highest power, energy, and area on-chip (Table \ref{tab:asic_res}). The post-layout energy numbers were almost 2 $\times$ higher than the post-synthesis results. This increase in energy is mainly because of metal routing resulting in layout parasitics. As the tool uses different routing optimizations, the power, area, and energy values keep changing with the classifiers' composition and architecture.
	\begin{table}[htb!]
		\centering
		\vspace{-0.75em}
		\caption{Post synthesis hardware results for proposed framework with different ML classifiers (@100MHz) }\label{tab:asic_res}
		\scalebox{0.8}{
			\begin{tabular}{|c|c|c|c|}
				\hline
				
				Classifier &  Power ($mW$) & Energy ($mJ$) & Area ($mm^2$) \\
				\hline 
				AlexNet & 72.45  & 5.22 & 4.5\\
				\hline
				ResNet-50 & 68.64  & 3.75 & 3.55 \\
				\hline
				MobileNet & 64.63  & 3.79  & 3.81  \\
				\hline
			    CNN & 46.44  & 2.29 & 2.45  \\
				\hline
				VGG-16 & 56.46 & 3.21 & 3.26 \\
				\hline
		        \end{tabular}}
		\vspace{-1.5em}
	\end{table}

\section{Conclusion}
\label{conclusion}
With the proposed code-aware generative AI technique we were able to employ few-shot learning to efficiently classify, limitedly seen malware data. From the results presented, it is evident that the model trained on data, generated through the proposed technique outperforms, the model trained on limited available data, with approximately 9\% more accuracy. We also furnished the ASIC implementations of different classifiers trained using the proposed technique. Thus, instead of training the ML models with limited available data, the proposed code-aware data generation technique should be employed.

\bibliographystyle{IEEEtran} 

\bibliography{ref.bib}

\end{document}